# A Bayesian approach to technological emergence, climate variability, and demographic stability in northern South America

Sebastian Fajardo[a,*], Philip Riris[b], María Vélez[c], Natalia Álvarez Correa[d,e], A. Jacobo Hidalgo[f], Luis Felipe Navarro[d], Rodolfo Cepeda[d], Nicolás Loaiza[g], Carlos Reina[d], Javier Ruiz-Pérez[h], Marco Madella[h,i,j]

[a] Leiden Institute of Advanced Computer Science (LIACS), Universiteit Leiden, Einsteinweg 55, 2333 CA Leiden, The Netherlands.

[b] Institute for the Modelling of Socio-Environmental Transitions, Bournemouth University, BH12 5BB Fern Barrow, United Kingdom.

[c] Department of Earth Sciences, University of Regina, 3737 Wascana Parkway, S4S 0A2 Regina, Canada.

[d] Instituto Colombiano de Antropología e Historia (ICANH), Calle 12 No. 2-41, 111711 Bogotá, Colombia.

[e] Departamento de Antropología, Universidad Nacional de Colombia, Carrera 30 # 45-05 111321 Bogotá, Colombia,

[f] Ingecontrol S.A., Calle 14 B No. 109-45, 110111 Bogotá, Colombia.

[g] Fundación Pro Calima Bogotá, Colombia

[h] Department of Humanities, CASEs group , Universitat Pompeu Fabra (UPF), carrer de la Mercé, 08002 Barcelona, Spain.

[i] ICREA, Pg. Lluís Companys 23, 08010 Barcelona, Spain.

[j] School of Geography, Archaeology, and Environmental Studies (GAES), University of the Witwatersrand, 1 Jan Smuts Avenue, Braamfontein 2000 Johannesburg, South Africa.

[*] Corresponding author. E-mail address: s.d.fajardo.bernal@liacs.leidenuniv.nl (S. Fajardo).

## Highlights

- Human-activity growth accelerated ~13-fold after a changepoint at 3600 cal BP.
- Plant management and pottery production predate the shift in human activity; metallurgy and raised fields cluster at its onset.
- Reduced El Niño suppressed pre-Columbian activity; stronger El Niño mildly raised it.
- Three independent ENSO proxies converge on the same asymmetric coupling pattern.
- Technological emergence behaves as a phase transition in collective computation, not stress.

## Abstract

Human communities produce technology in almost every environment, but the conditions that enable technologies in the tropics are not well understood. We test two competing hypotheses—that technological emergence is triggered by environmental stress and that technology tracks growth in effective populations—by integrating Holocene archaeological and paleoclimate evidence. We use archaeological derived radiocarbon dates from Colombia as a proxy for human activity, including dating plant management, pottery, metallurgy, and raised fields. For palaeoclimate, we use two El Niño–Southern Oscillation (ENSO) reconstructions: Laguna Pallcacocha and Laguna El Junco in Ecuador. A Bayesian changepoint at 3575 cal BP (95% highest posterior density interval (HPDI) = 3811–3329) marks a ~13-fold increase in the growth rate of pre-Hispanic human activity ($r_2/r_1$ median = 13.3, 95% HPDI 8.1 - 22.2). Plant management and pottery predate this shift by more than a millennium, whereas metallurgy and raised fields cluster within the early acceleration phase, during relative climatic and demographic stability. A Bayesian log-linear regression of human activity on ENSO frequency and intensity after the demographic changepoint reveals an asymmetric covariation: reduced ENSO frequency and intensity suppress activity by 58% and 21%, whereas ENSO frequency raises activity modestly (~11-13%). Therefore, ENSO regulates demographic activity rather than triggering technological emergence. These patterns suggest that pre-Columbian communities may have innovated when they could compute—when an effective population sat above a critical threshold and stability windows permitted combinatorial search—rather than when they were forced to by environmental stress, expressed as demographic or climatic variability.

## 1. Introduction

Understanding why and how technologies emerge, spread, and persist is a central challenge in archaeology. This issue has been investigated using evolutionary and formal models, commonly described as cultural evolution (Creanza et al., 2017a; Mesoudi, 2011), alongside ecological (Baumard and André, 2025), action-centred (Ingold, 2016; Pauketat, 2001), and cultural-historical approaches (Trigger, 1989). Most evaluations of cultural evolution models in archaeology rely on data from high-latitude regions like Western and Central Europe (Matzig et al., 2023). These regions feature prominently because they have excellent chronological sequences, abundant and durable material culture (pottery, tools, metal objects), and large-scale datasets suitable for statistical modelling. Cultural evolutionary archaeology in the tropics has largely focused on social complexity (Drennan and Peterson, 2006; Hamilton et al., 2020; Matzig et al., 2023; Turchin et al., 2022) and on food resource adaptations (Clement et al., 2021; Fort, 2025; Furquim et al., 2021; Iriarte et al., 2020; Lombardo et al., 2025; Souza et al., 2025, 2020). While some studies used ethnographic evidence from small-scale societies in tropical regions to explore transmission networks (Henrich and Broesch, 2011; Jang et al., 2024; Schillinger and Lycett, 2019) and segregation and succession in transmission dynamics (O'Brien et al., 2016), most cultural evolution models remain largely untested in tropical archaeological contexts. To address this gap, we focus on the Colombian Holocene as a case study to document which environmental and demographic patterns best explain the emergence of plant management, pottery, metallurgy, and raised fields, framing the emergence of these technologies as a collective computation problem.

## 1.1. Climate stress and effective population shape innovation processes

Rates of inventions and innovations in human societies are shaped by environmental factors; communities may generate innovations in response to the stress or abundance these factors create. Climate variability and climate change can act as selective pressures as populations confronted with changing conditions may invent new technologies or behaviours (Aoki et al., 2005; Collard et al., 2013; Morgan et al., 2022; Rendell et al., 2010; Richerson et al., 2001). Here, we treat interannual hydroclimatic variability as a form of environmental stress. For populations whose food production depends heavily on the predictability and consistency of weather patterns, such variability can force the continual reconfiguration of subsistence technologies. When environmental conditions are stable, by contrast, innovating via individual learning becomes costly, and communities would favour their accumulated knowledge, rather than favouring innovation (Morgan et al., 2022; Perreault et al., 2012).

Effective populations also shape the innovation process (Deffner et al., 2022; Powell et al., 2009; Shennan, 2001). The rate at which novel cultural variants accumulate and spread is a function of the number, density, and connectivity of interacting agents capable of transmitting and recombining ideas (Derex et al., 2018). In its modern formulation, innovation is best understood as driven by the effective size of the population exchanging ideas rather than as the product of individual derived creativity (Bentley and O'Brien, 2024). Here we treat demographic variability as another form of environmental stress that can destabilize effective populations capable of sharing or creating ideas. Effective population size—a function of mobility, social network structure, raw population size, density, and intensity of interaction—shapes the expected rate of innovation through two channels. First, even when the per-capita innovation rate is held approximately constant, a larger effective population may produce more innovation per unit time simply because more individuals contribute. Second, effective population size can raise the per-capita rate in larger, denser, and more connected populations because individuals have access to more ideas to recombine (Creanza et al., 2017b; Kolodny et al., 2016).

Climate variability and effective population size matter, but recent studies show their effects are neither deterministic nor easily disentangled, and need to be evaluated jointly (Burke et al., 2025). Strong selection can push populations toward social learning even under unstable environments, because the fitness cost of getting a technology wrong, outweighs the value of experimentation (Premo and Kuhn, 2010). Among small populations, cultural drift can fix or eliminate neutral traits quickly producing patterns that mimic biased social transmission or selection. Population size may thus act less as a direct driver of innovation and rather as a modulator of stochasticity (Carrignon et al., 2024; Premo, 2016). Path dependence adds further inertia: the learning strategies with which a population starts, persist even when ecological conditions would predict a shift (Demps et al., 2024). Moderate environmental variability produces oscillation between strategies rather than a clean signal, so the same underlying dynamic can leave divergent archaeological expressions across regions (Deffner et al., 2020). Innovation, then, emerges from the joint action of selection pressure, drift, starting conditions, and ecology. Although environmental stress and population characteristics are typically framed as very important in models for adaptation (Wren et al., 2025), few studies have evaluated them jointly against the same empirical record within a common inferential framework in tropical regions. This is the objective of the current work for the region of Colombia during the Holocene.

## 1.2. Technological emergence as collective computation

We frame the relationship between system variability and technological innovation as a problem in collective computation (Flack, 2017; Krakauer, 2011), in which a population of interacting agents generates novelty by recombining existing cultural elements (Arthur, 2009; Kauffman, 1993; Weitzman, 1998). The model couples two distinct mechanisms. The first mechanism is a demographic prerequisite: sustained innovation requires that the effective population, together with the diversity of transmissible ideas, exceeds a threshold below which cumulative culture cannot be sustained. Below this threshold, novel variants (inventions) are not necessarily lost. Traits can persist by either of two routes: they may diffuse at a slow pace from a single originator or they may arise repeatedly and independently across communities emerging by convergence rather than by transmission. Adoption strung out across long intervals rather than taken up across the population more or less at once. Technological clusters in such scenarios are correspondingly rare. Such a dynamic is hypothesised for the Tasmanian Holocene technological loss (Henrich, 2004), and it is consistent with broader cross-cultural and experimental evidence that effective population size constrains cultural complexity (Derex et al., 2013; Derex and Mesoudi, 2020;

Kline and Boyd, 2010). It is also used to explain the late appearance of behavioural modernity (Powell et al., 2009). This mechanism should show a changepoint, more specifically, a percolation-like or autocatalytic transition (Kauffman, 2003, 1993): once the network of recombining ideas crosses this threshold, it begins to feed itself—each innovation enlarges the pool from which the next one is drawn, and the process becomes self-sustaining. The second mechanism operates only above this threshold. Even within the supercritical regime, the rate and magnitude of innovation depend on the system's proximity to a critical point at which susceptibility to perturbation is high and system shifts can trigger large enough reorganisations of the cultural network to be traced in the archaeological record (Potts, 2013; Richerson et al., 2001). This is the divergent-susceptibility property characteristic of critical systems (Mora and Bialek, 2011; Muñoz, 2018). Climate or demographic variability act as control parameters on this second mechanism: high-amplitude or high-frequency fluctuations destabilise the conditions to recombine technological solutions and reduce time to find adjacent regions of the design space. The two mechanisms are empirically separable: demographic conditions control whether a cluster of co-occurring technologies is *possible*, while climate and demographic variability conditions modulate its *tempo*. The emergence of clusters of co-occurring technologies is therefore not a single phase transition but the joint product of a supercritical recombination process and an environmentally-tuned susceptibility. This framing (Figure 1) generates predictions that are different from those derived from purely demographic or purely climatic models.

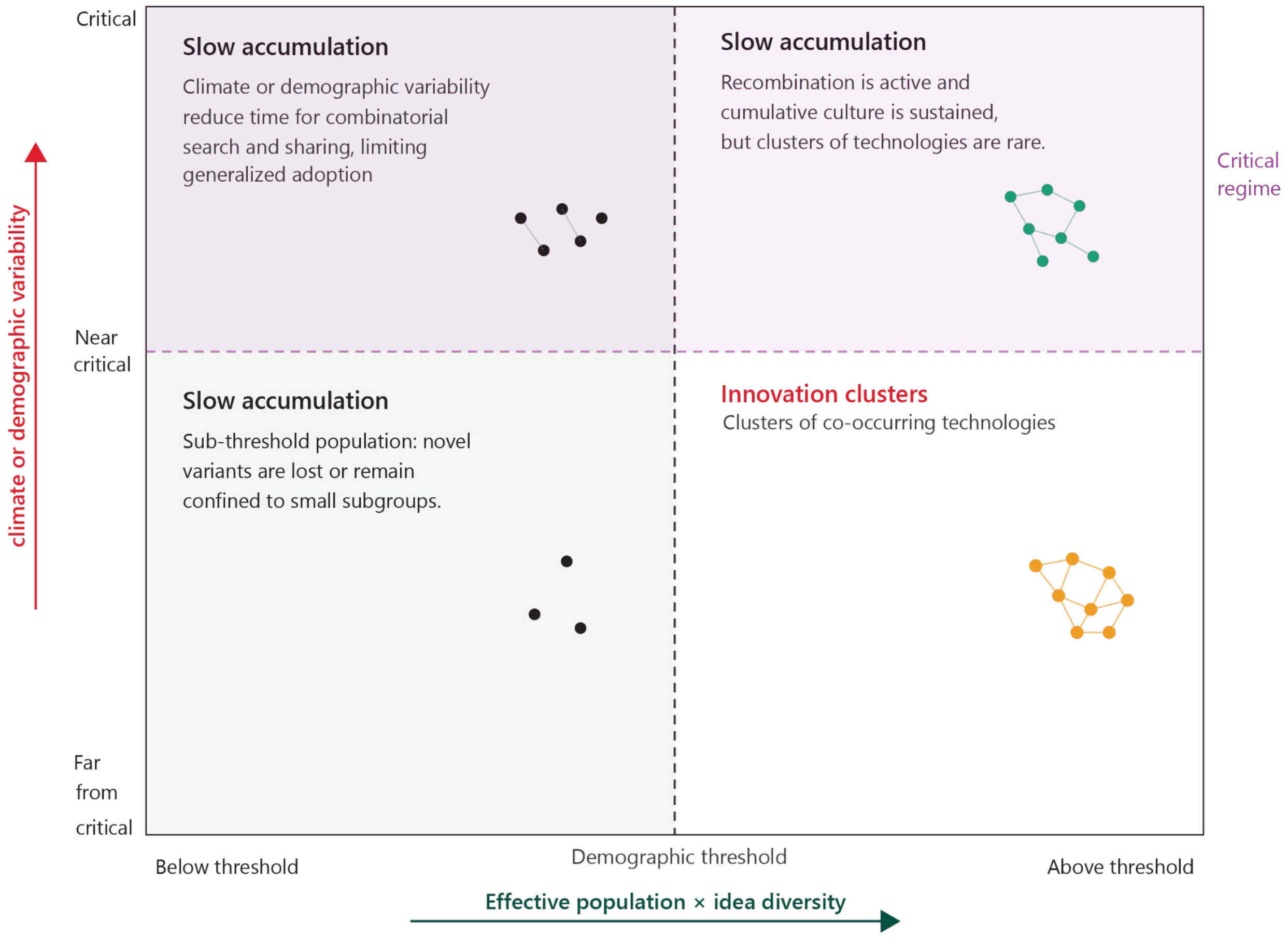

**Figure 1. Predicted regimes of technological emergence as a function of effective population and climate/demographic variability**. Innovation clusters arise only above the demographic threshold under comparatively stable conditions (lower right); the remaining quadrants produce slow accumulation.

We use this framing to propose a testable prediction. If the demographic threshold for an effective population is crossed under stable climate and demographic conditions, we expect the emergence of innovation clusters—defined here as co-occurring technologies packed into a narrow window. Conversely, if the population threshold is crossed but conditions remain volatile, recombination and cumulative cultural evolution should persist, yet clusters of co-occurring technologies will likely be rare. Below the demographic threshold, regardless of environmental conditions, novel variants may be lost or remain confined to small subgroups, and when widespread adoption occurs, it happens over long intervals of time, and clusters or technologies are unlikely to form.

We confront these predictions with the archaeological evidence from northern South America, using data from Colombia and environmental proxies from Ecuador. Present-day Colombia provides a case study with innovations spanning three knowledge domains: the management of living systems (plants), the modification of material properties (pottery and metallurgy), and the modification of water and terrain (raised fields). These innovations arose within precolumbian societies with dispersed households, few or no nucleated settlements, and low hierarchical organization (Drennan et al., 2026), letting us examine the demographic and environmental drivers of cumulative cultural evolution under conditions of low daily face-to-face interaction. We pair the Andean (Pallcacocha) and equatorial-Pacific (El Junco) proxies because no Colombian site offers a continuous, high-resolution, and open-access El Niño–Southern Oscillation (ENSO) archive independent of pollen-based reconstructions. The influence of ENSO on Colombian climate variability is well established (Poveda et al., 2011), operating through teleconnections, so we take the variability registered in Ecuador as a proxy for the variability experienced in Colombia. Nevertheless, ENSO's terrestrial expression in Colombia is spatially heterogeneous, with the sign and magnitude of precipitation anomalies differing across the Caribbean coast, the bimodal-rainfall inter-Andean valleys, and the Amazonian piedmont (Velez et al., 2026).

Our analytical framework integrates three components: (i) a double-exponential Bayesian growth model fit to the Colombian radiocarbon record, which identifies the timing of the demographic threshold and demographic stability windows; (ii) a Bayesian log-linear regression of post-shift activity intensity on climate using ENSO frequency and intensity, as proxy for environmental variability; and (iii) a formal alignment of the inferred changepoint, the climate deviations, and the posterior intervals for the earliest documentation of plant management, pottery, metallurgy, and raised field structures.

# 2. Materials and methods

## 2.1. The Colombian Holocene technological record as a test case

We focus on the Holocene archaeological record from Colombia and examine the emergence of plant management, pottery, metallurgy, and raised fields. We treat these as transformative technologies, that is, technologies that alter or control properties of elements in the environment by drawing on distinct domains of knowledge.

The earliest directly dated evidence for plant management and cultivation in Colombia comes from preceramic sites of the middle Cauca and the Eastern Cordillera, where plant remains and processing tools were dated by stratigraphic association with carbonized remains. The oldest date comes from El Jazmín, in the middle Cauca, at 11702 cal BP (10120 ± 70 BP, Ua-24497), in deposits with grinding stones, handstones and unifacial tools alongside macro- and microbotanical remains of managed palms (*Aiphanes horrida* (Jacq.) Burret) and tubers (*Dioscorea* sp. - yam, *Xanthosoma* sp. - cocoyam) (Aceituno and Loaiza, 2007). Comparable early evidence comes from La Morena, in the Aburrá Valley, at 11583 cal BP (10060 ± 60 BP, Beta-245566), with yam (*Dioscorea* sp.), beans (*Phaseolus* sp.) and palm remains together with a ground-stone toolkit (Santos, 2010), and from San Isidro, in Popayán Plateau, at 11574 cal BP (10050 ± 100 BP, Beta-65878), where edge-ground cobbles, a polished axe and remains of avocado (*Persea americana* Mill.), mompás/chachafruto (*Erythrina edulis* Triana ex Micheli) and calabazo (*Lagenaria siceraria* (Molina) Standl.) were found (Gnecco Valencia and Bravo, 1995). A direct date on plant remains from other sites yielded younger chronologies: charred seeds from diverse taxa palms, phytoliths of squash (*Curcubita* spp.), *L. siceraria*, and bijao (*Calathea* sp.) from Peña Roja (Morcote-Ríos et al., 2014), in the Amazonian piedmont were dated at 9824 cal BP (8800 ± 40 BP, Beta-332853). Given the composition of the plant assemblages and the fact that the oldest radiocarbon determinations were obtained from charred wood rather than the cultigens themselves, we interpret these data as evidence for the onset of managed-plant assemblages. The origins of plant domestication in present-day Colombia reflects long-term processes that unfolded over a diversity of geographies (see for a review Aceituno, 2026). Species such as maize and common bean reached Colombia already domesticated from Mesoamerica and the Central Andes (Aceituno and Loaiza, 2014; Pearsall, 2008). However, the tropical lowlands of northern South America have been recognised as one of the independent centers of domestication of tubers (e.g. manioc, arrowroot, leren, sweet potato), squashes, and tree fruits (e.g. peach palm) during the early and middle Holocene (Aceituno, 2026; Aceituno and Loaiza, 2018; Lombardo et al., 2020; Piperno, 2011; Souza et al., 2025).

The earliest pottery in Colombia, and among the oldest in the Americas, comes from the San Jacinto and lower Magdalena region of the Caribbean lowlands. The earliest radiocarbon dates for pottery-bearing contexts, derived from charred wood, originates from San Jacinto I, in the north of Colombia, at around 6769 cal BP (5940 ± 60 BP, PITT-0155), associated with fibre-tempered ceramics in an incipient-sedentary, plant-processing context (Oyuela-Caycedo and Bonzani, 2005). Shell-midden sites extend the record into the late sixth millennium: Puerto Chacho (Legros, 1992) at 5998 cal BP (5220 ± 90 BP, Beta-26200, charred wood), Puerto Hormiga (Reichel Dolmatoff, 1965) at 5536 cal BP (4820 ± 100 BP, SI-151, charred wood) to 5169 cal BP (5040 ± 70 BP, SI-153, shell), and Monsú (Reichel-Dolmatoff, 1985), at around 5472 cal BP (5300 ± 80 BP, UCLA-2149C, shell). Early pottery function in northern South America may have depended on the type of temper employed, with pottery tempered with vegetable fibers used primarily for transport, storage of water or food, and serving (Pratt, 1999), whereas pottery tempered with mineral was used for cooking-related activities (Carvajal Contreras, 2022). In the Americas, other early pottery centers are documented in the Amazon at the shell midden of Taperinha, in present-day Brazil. Twelve radiocarbon dates on shell, charred wood, and, in one case, carbonized residue from a ceramic sherd, and a single thermoluminescence date, place the pottery production at around 7000 cal BP (Roosevelt et al., 1991).

The earliest directly dated evidence for goldworking comes from the Pacific coast at Tumaco-La Tolita, and the Caribbean lowlands of the lower San Jorge River. The oldest date is on the clay-and-charcoal casting core recovered inside a cast-gold cane finial in the form of a bicephalous deer from Majagual (Sucre), at 3539 cal BP (3320 ± 50 BP, Beta-82928), directly dating a lost-wax casting event (Plazas, 1998). A second piece from the same study, a gold pendant from San Marcos (Sucre), dates to 3013 cal BP (2880 ± 60 BP, Beta-105880); both have only estimated provenience, so should be treated with some caution. More securely contextualised evidence comes from the Pacific coast at Tumaco: hammered gold sheets with repoussé decoration and goldsmithing debris from the artificial mound TR-2 at La Magnolia, dated on charred wood to 2502 cal BP (2420 ± 90 BP, Beta-82931) (Patiño, 2003, 1997; Plazas, 1998), and hammered gold wires from Inguapi at 2258 cal BP (2275 ± 85 BP, Ny-642) (Bouchard, 1984). The earliest metallurgy evidence from Colombia, whether from the San Jorge River area or from the Pacific coast at Tumaco-La Tolita was embedded in a wider technological circulation. The earliest known archaeological evidence of metalworking in the Americas dates between 4105 and 3886 cal BP **(**3733 ± 43 BP, AA-36815**)**, and it is associated with cylindrical beads of cold-hammered native gold found with a human burial at Jiskairumoko in the Titicaca Basin (Aldenderfer et al., 2008; Zori, 2019). Taken together, this evidence points to northern South American contacts and influences from the Andean centers to the south between the third and second millennia before present (Lleras Pérez, 2007). This indicates that these innovations did not develop in isolation but probably as part of a larger regional interaction network.

The earliest directly dated evidence for raised fields and channelled fields comes from the Sabana de Bogotá and the Caribbean savannas. The oldest date (3260 cal BP / 3050 ± 40 BP, GX-30239) is on humic acid from a soil sample associated with maize pollen at Guaimaral, in the Sabana de Bogotá (Boada, 2018). A second humic-acid date from La Filomena was dated 2523 cal BP (2450 ± 40 BP, GX-30234) in a comparable maize-bearing context. In the lower San Jorge floodplain, charred wood from the large Zenú hydraulic systems dates the canal-and-platform networks at 2892 cal BP at Pimienta 5 (2760 ± 120 BP, GrN-14472) and 2262 cal BP at Carate 19 (2280 ± 80 BP, GrN-14475), both part of a canal complex extending some 60 km over a palaeochannel of the San Jorge River (Plazas et al., 1993). Raised fields and canals emerged as a response to water management in flood-prone landscapes such as the Sabana de Bogotá, the lower San Jorge basin, and the Gulf of Urabá, allowing communities to intensify agricultural production and sustain dense populations in alluvial valleys (Boada, 2018; Gallo, 2019; Plazas et al., 1993; Posada-Restrepo et al., 2023). Raised fields are common in other regions, such as the Llanos de Moxos, where OSL dating (sample FFM, 1340 ± 100) indicates the fields were in use as early as 570–770 BP (Rodrigues et al., 2015). Other human-made landscape modifications and water management in South America have older dates such as in the Zaña Valley of northern Peru, where a sequence of buried irrigation canals has the following chronology: Canal 3 dates at 6310–5940 cal BP (5380 ± 80 BP, Beta-154127), and a possible fourth, deeper canal at 8040–7340 cal yr BP (6705 ± 75 BP, Beta-34332) (Dillehay et al., 2005)**.**

## 2.2. Radiocarbon dataset

As part of the project Mapping the Archaeological Pre-Columbian Heritage of South America (MAPHSA), we compiled COL-14, a dataset with 2083 pre-Columbian archaeological radiocarbon dates covering the present-day Colombia territory (Figure 2). We adopted the framework "dates as data" (Rick, 1987; Shennan and Edinborough, 2007), treating the assemblage as a proxy for the intensity of human activity rather than as a direct measure of past population size (Carleton and Groucutt, 2021; Williams, 2012). Such activity may reflect demographic growth, expanding interaction networks, more intensive work, frequent site use or human interaction, or some combination of these factors—distinctions that radiocarbon dates alone cannot resolve (Contreras et al., 2025). Dates were calibrated using IntCal20 for terrestrial samples and Marine20 for marine samples (Heaton et al., 2020; Reimer et al., 2020). All marine shell samples (n=25) derived from the Caribbean coast and were calibrated without an additional local reservoir correction, given the lack of ΔR estimates currently available for the Colombian Caribbean coast (DiNapoli et al., 2021). A summed probability distribution (SPD) was computed as a baseline using the *rcarbon* R package (Crema and Bevan, 2021), following established procedures to identify departures from a given null model (Shennan et al., 2013; Timpson et al., 2014). Combined laboratory and calibration-curve uncertainty was propagated through to the SPD. From the same dataset we isolated 70 dates associated with the earliest documented occurrences for the four transformative technologies: plant management (n = 35), pottery (n = 24), metallurgy (n = 7), and raised field agriculture (n = 4).

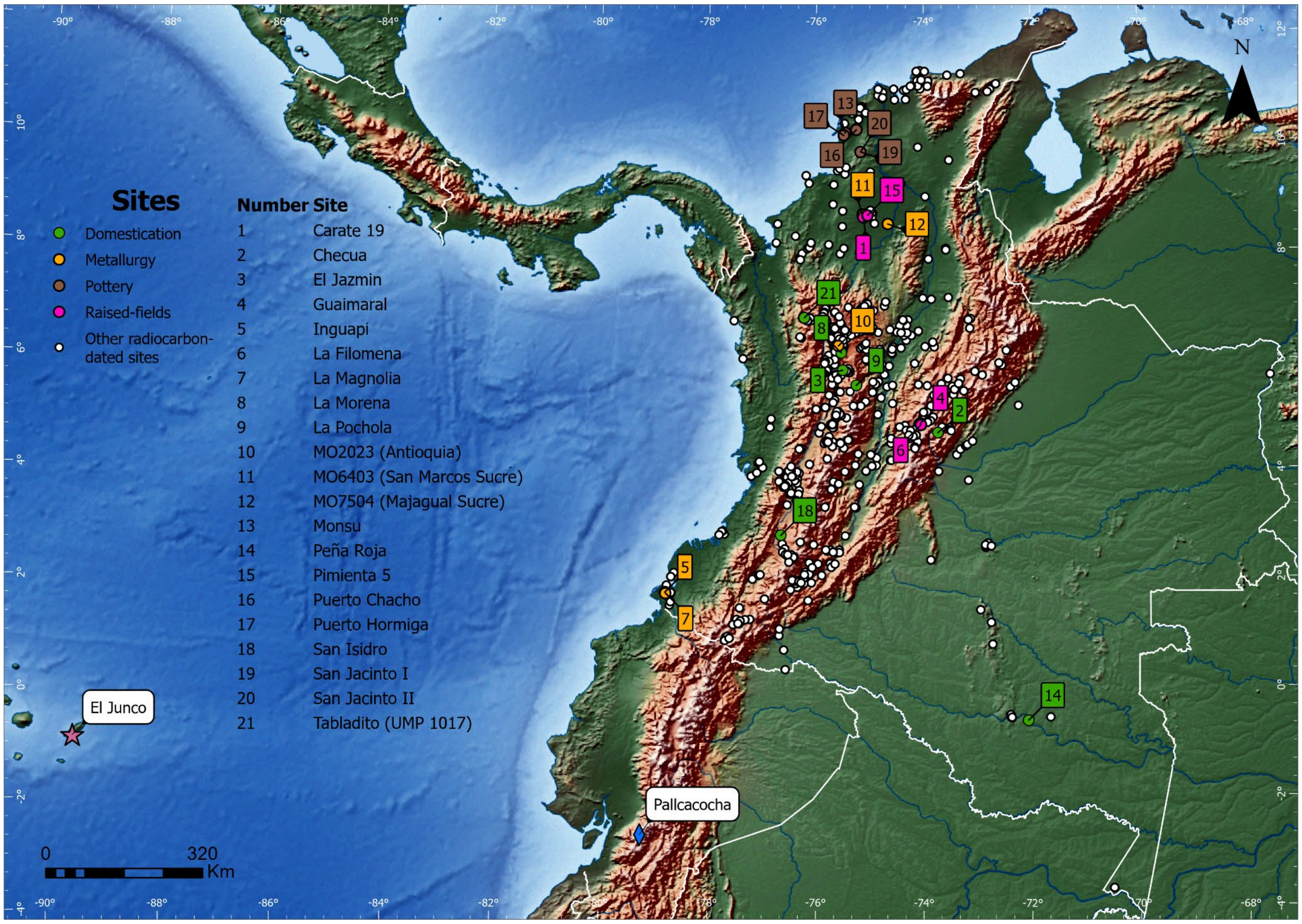

**Figure 2. COL-14 radiocarbon dates and archaeological and palaeoenvironmental sites discussed in this study**. Coloured square markers denote 21 approximate locations of archaeological sites and findings associated with early emergence of plant management (green; *2* Checua, *3* El Jazmin, *8* La Morena, *9* La Pochola, *14* Peña Roja, *18* San Isidro, *21* Tabladito [UMP 1017]), pottery (brown; *13* Monsú, *16* Puerto Chacho, *17* Puerto Hormiga, *19* San Jacinto I, *20* San Jacinto II), metallurgy (orange; *5* Inguapi, *7* La Magnolia, *10* MO2023 [Antioquia], *11* MO6403 [San Marcos, Sucre], *12* MO7504 [Majagual, Sucre]), and raised fields (magenta; *1* Carate 19, *4* Guaimaral, *6* La Filomena, *15* Pimienta 5). Small white circles mark additional radiocarbon-dated archaeological sites with known approximate coordinates (n=1578). The pink star and blue diamond mark the location of palaeoenvironmental records of El Junco, in the Galápagos Islands, and Pallcacocha, in the southern Ecuadorian Andes, respectively.

## 2.3. Environmental proxies

Reconstructing past climate in Colombia is inherently complex (Mesa et al., 2021), as the three Andean cordilleras and the combined influence of the Pacific, Caribbean, and Amazon Basin generate a mosaic of microclimates that no single proxy can fully capture (Velez et al., 2026, 2021). Despite the multifactorial nature of Holocene climate dynamics in Colombia, ENSO was selected as the principal climatic framework because it constitutes a primary source of interannual hydroclimatic variability in the region. Changes in ENSO state strongly regulate precipitation, seasonality, river discharge, and drought frequency, with El Niño phases generally associated with widespread moisture deficits across Colombia (Poveda et al., 2011). Importantly, ENSO integrates large-scale ocean-atmosphere interactions whose impacts are expressed throughout northern South America, allowing local paleoecological records to be interpreted within a coherent regional context. Consequently, ENSO provides the most robust and regionally meaningful large-scale proxy currently available for investigating Holocene environmental variability in Colombia.

We used three published time series for the ENSO conditions. First, we used XRF-PC1, a record derived from a principal component analysis of the sedimentary sequence from Laguna Pallcacocha (Ecuador), which is interpreted as a proxy for El Niño event frequency. Higher values indicate greater El Niño frequency (Mark et al., 2022). Second, we used the botryococcene δD record from Laguna El Junco (Galápagos), interpreted as a proxy for the intensity (amplitude) of El Niño conditions, where lower δD values indicate greater amplitude (Zhang et al., 2014). Third, we used the log-concentration botryococcene record from El Junco, interpreted as a frequency proxy for ENSO conditions, where higher values indicate more frequent El Niño events (Zhang et al., 2014). All proxies were detrended, re-binned to 100-yr resolution, and standardised using z-scores prior to inclusion in the climate–activity regression (see data availability). We selected the Pallcacocha and El Junco records as regional—rather than local—hydroclimate proxies because they offer the highest resolution and are the most proximal openly accessible records to Colombia. Although speleothem records from the region exist (Muñoz et al., 2017), the underlying data are not openly available, so the Pallcacocha and El Junco series remain the best proxies for our analysis.

## 2.4. Bayesian double-exponential growth and changepoint model

To identify the timing of the demographic shift we fitted a double-exponential growth model to the radiocarbon dates within a Bayesian framework. The model treats the calendar age of each date as a latent parameter whose prior is the growth model. Let *t* denote calibrated years before present and *p(t)* the modelled probability mass at time *t*. The deterministic part takes the form of Eq. (1), with *a* = 12900 and *b* = 500 cal BP the older and younger bounds of the analytical window,

$$p(t) \propto \begin{cases} \exp\big(-r_1\,(t-a)\big), & chp \le t \le a, \\ \exp\big(-r_1\,(chp-a)\big)\ \exp\big(-r_2\,(t-chp)\big), & b \le t < chp, \end{cases} \quad (1)$$

where $chp$ is the changepoint and $r_1$ and $r_2$ are the per-year proportional growth rates of the older (pre-changepoint, $t \geq chp$) and younger (post-changepoint, $t < chp$) segments, respectively. Here $p(t)$ is a categorical probability mass function over the interval 12900–500 cal BP. Therefore, the model assumes the rate at which dated events accumulate changed at a single point in time, and it estimates three quantities from the data: when that changing point occurred ($chp$) and how fast events were accumulating before it ($r_1$) and after it ($r_2$). We placed weakly informative, non-negative priors on the rates, $r_1$, $r_2$ ~Exp(mean=0.0004), based on the average growth rate reported across multiple prehistoric populations (Zahid et al., 2016) and used as a prior in other archaeologically oriented Bayesian models (Crema and Shoda, 2021). The changepoint was given a normal prior centred on 3500 cal BP, with a standard deviation of 200 years, truncated to the limits of the calibrated chronology (12900–500 cal BP). We used for the inference the *nimbleCarbon* package (Crema and Shoda, 2021), with four Markov chain Monte Carlo (MCMC) chains of 100000 iterations each, a burn-in of 10000, and a thinning interval of two, yielding 45000 post-burn-in samples per chain.

## 2.5. ENSO modulation of activity intensity

To test whether ENSO modulates human activity after the demographic threshold, we fitted a Bayesian Poisson-like regression of binned SPD fractional counts on standardised ENSO predictors within the post-changepoint window. For each 100 year bin $t$ with before present age as $age_t$,

$$\log \lambda_t = \alpha + r\,(a - age_t) + \beta_+ \cdot ENSO_t^+ + \beta_- \cdot ENSO_t^-, \ (2)$$

$$\lambda_t = \exp(\log \lambda_t),\ n_t \sim Normal(\lambda_t, \sqrt{\lambda_t}), \ (3)$$

where $a$ is the older bound of the analysis window, taken from the upper 95% highest posterior density interval (HPDI) of the changing point of the double-exponential model and $r(a - age_t)$ removes the underlying demographic trend so that the $\beta$ coefficients capture climate effects per bin. Because the model is specified on the log scale, each $\beta$ is a multiplicative effect on the expected event rate per +1 SD change in the predictor; the corresponding percentage change in human activity is $(e^{\beta} - 1) \times 100$. For the XRF-PC1, the predictors are rectified, strictly non-negative deviations from zero, $ENSO_t^+ = max(0, z_t)$ and $ENSO_t^- = |min(0, z_t)|$, where $z_t$ is the standardised proxy value at bin $t$. Under this, $\beta_+$ and $\beta_-$ act as one-way valves: a negative $\beta_-$ can only suppress $\log \lambda_t$ during reduced-ENSO intervals, and a positive $\beta_+$ can only enhance it during enhanced-ENSO intervals.

For the El Junco proxies (log-Botryococcene concentration and $\delta D$), we instead use signed two-way z-scores, so that a single $\beta$ coefficient simultaneously encodes enhancement under one sign and suppression under the other, yielding a symmetric slope. For each bin $t$,

$$\log \lambda_t = \alpha + r\,(a - \mathrm{age}_t) + \beta \cdot ENSO_t, \quad (4)$$

$$\lambda_t = \exp(\log \lambda_t),\; n_t \sim Normal(\lambda_t, \sqrt{\lambda_t}), \quad (5)$$

where $\mathrm{ENSO}_t$ is the signed standarised proxy value at bin $t$.

Priors were $\alpha \sim$ Normal ($\mu = 0$, $\sigma = 10$), $r \sim$ Exp(1/0.0004), and $\beta$, $\beta_+$, $\beta_- \sim$ Normal ($\mu = 0$, $\sigma = 1$). Sampling used four nimble chains of 100000 iterations each with a 10000 burn-in and a thinning interval of two.

## 2.6. Bayesian estimation of technology phase boundaries

For each technology we estimated the start and end of its dated phase within a Bayesian framework using OxCal v4.4 (Bronk Ramsey, 2009). Each technology was modelled as a single phase of uniformly distributed events bracketed by a start and an end boundary, and the four technologies were treated as independent sequences. We treated these boundaries as the emergence window of each technology. The boundaries are estimated parameters rather than observed dates. Each boundary is summarised by its posterior median and 95.4% highest-posterior-density interval in cal BP, and the duration of each phase as the difference between the start and end boundaries; these are reported in Section 3.2 and Table 1.

## 2.7. Climate variability, human activity and technology emergence alignment

We placed the archaeological and paleoenvironmental evidence in a common temporal frame to compare: (i) the posterior distribution of the changepoint in the intensity of human activity, estimated from the double-exponential growth model; (ii) the deviations of the observed SPD from the model's posterior predictive envelope, which we treat as demographic departures from the expected trajectory; and (iii) the posterior intervals for the earliest documented occurrence of each technology (plant management, pottery, metallurgy, and raised field construction). We then estimated the strength with which climate variability modulates activity intensity.

We restrict the climate-modulation analysis to the interval after the changepoint (see section 2.4). Before the changepoint, the regional population appears to have been small, dispersed, and reliant on broad-spectrum foraging rather than on plant management/cultivation as its principal subsistence base (Aceituno, 2026; Delgado Burbano, 2012). Short-term climate fluctuations do not appear to have left a strong demographic signal; the diet breadth and residential mobility of these groups would have buffered them against the climate-driven shortfalls to which later, more cultivation-dependent and aggregated populations were exposed (Halstead and O'Shea, 1989; Riris et al., 2024; Wren et al., 2025). Concentrating the analysis on the post-changepoint window therefore tests the modulation hypothesis where the proposed mechanism should operate.

Against this backdrop, we asked three questions. First, does the posterior intervals for the technologies cluster in time, and whether that cluster falls before or after the posterior changepoint in activity intensity? Second, do these intervals coincide with periods of high climate variability in the proxy records or with demographic departures—intervals in which the observed SPD lies outside the growth-model envelope—or do they instead fall within windows in which both climate and demography are comparatively stable? Third, how much of the post-changepoint demographic departure is attributable to climate variability? We quantify this by estimating, in a log-linear regression of the SPD-derived bin counts, the marginal effect of each climate proxy on binned activity intensity.

Analyses were performed in (R Core Team, 2026) using the *nimble* (Valpine et al., 2017), *nimbleCarbon* (Crema and Shoda, 2021), and *rcarbon* (Crema and Bevan, 2021) packages. Technology windows were estimated with a Bayesian framework using OxCal v4.4 (Bronk Ramsey, 2009). Model code, posterior samples, convergence diagnostics, and a snapshot of the dataset are deposited in a public repository (Fajardo et al., 2026 ; see data availability section).

# 3. Results

## 3.1. A demographic shift at ~3600 cal BP

The double-exponential growth model convergence was assessed with the Gelman–Rubin statistic and effective sample size (ESS) across four chains (180000 post–burn-in draws total). All parameters mixed well, with R-hat ≈ 1.00 and ESS exceeding 12000 for every parameter ($r_1$ = 20674; $r_2$ = 17521; changepoint = 12420). We estimate the demographic threshold at a posterior median of 3575 cal BP (95% HPDI 3811-3329 cal BP). Before this changepoint, the growth rate is essentially flat ($r_1 \approx 7.7 \times 10^{-5}$ yr$^{-1}$; 95% HPDI $4.1 \times 10^{-5}$ to $1.1 \times 10^{-4}$), equivalent to only about 0.8% population growth per century and consistent with roughly 9325 years of demographic quasi-stability between approximately 12900 and 3575 cal BP. The post-changepoint rate is more than an order of magnitude larger ($r_2 \approx 1.0 \times 10^{-3}$ yr$^{-1}$; 95% HPDI $9.5 \times 10^{-4}$ to $1.1 \times 10^{-3}$), corresponding to roughly 11% growth per century, and yielding an acceleration ratio $r_2/r_1 \approx 13.3$ (95% HPDI 8.1 - 22.2) (Figure 3).

## 3.2. Technology intervals relative to the demographic curve

Aligning the directly dated technology occurrences with the inferred demographic curve reveals a pattern consistent with the expectation of technological emergence as a collective-computation problem (Figure 3, Table 1). Convergence was monitored through the MCMC chains and model fit assessed with the OxCal agreement indices, with individual indices inspected for outliers (overall Amodel = 98.8, Aoverall = 99.2). no radiocarbon determination fell below A = 60 (minimum A = 88.8), indicating no significant outliers Boundaries are inferred rather than read off the extreme dates, and their intervals widen for technologies represented by fewer determinations (n = 7 metallurgy, n = 4 raised fields). The earliest dated evidence (start) yields median emergence ages of 11746 cal BP for plant management, 6780 cal BP for pottery, 3663 cal BP for metallurgy, and 3471 cal BP for raised fields. The latest dated evidence (end) yields medians of 7545, 4761, 2058, and 2085 cal BP, respectively.

Plant management and pottery emergence and early diffusion window occurred well before the demographic shift, and their intervals fall outside the 95% HPDI of the changepoint. The two do not overlap, indeed the latest estimated end median of the early diffusion window for plant management (7604 cal BP) predates the start median estimate for the emergence of pottery (6813 cal BP) by roughly 800 years. By contrast, the emergence of metallurgy and raised fields falls within the changepoint's 95% HPDI: the two start medians overlap with each other and with the changepoint interval, with metallurgy emerging roughly at the posterior median changepoint (3575 cal BP) and raised fields a few centuries later.

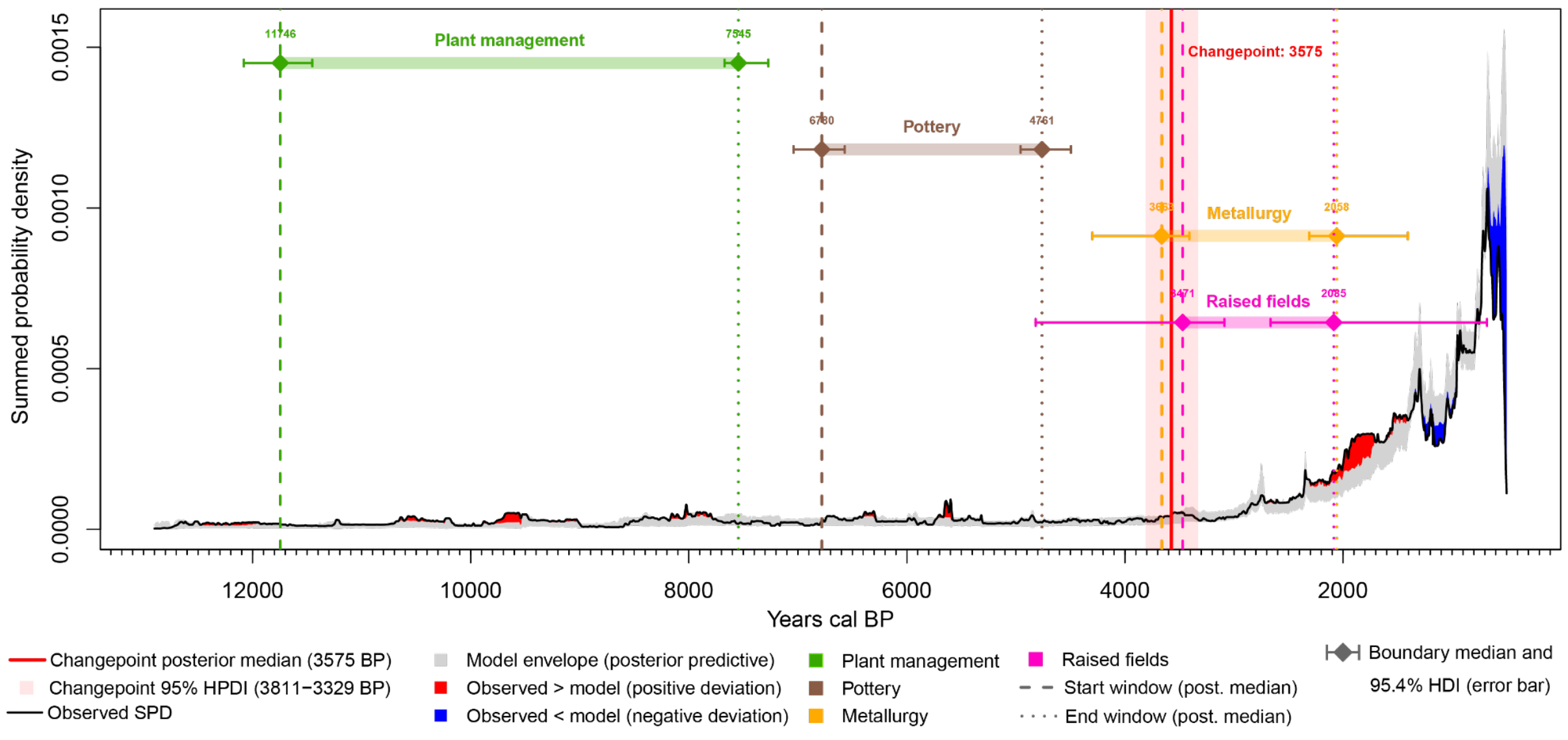

**Figure 3. Technology intervals overlaid on the posterior predictive check of the double-exponential growth model fitted to the radiocarbon record.** The observed SPD (black line) is plotted against the model's posterior predictive envelope (grey shade), generated by simulating SPDs (uncalibrated-sampling method; simulations = 1000) from the fitted model. Parameters were drawn from the joint posterior the of two growth rates $r_1$ (before changepoint) and $r_2$ (after changepoint). Intervals where the observed SPD rises above the envelope (more dated events than

the model predicts) are shaded red; intervals where it falls below the envelope (fewer than predicted) are shaded blue. The solid red vertical line marks the posterior median of the demographic changepoint (3575 cal BP) and the translucent red band shows its 95% HPDI (3811-3329 cal BP). For each technology, the start and end of the phase are represented as bars where filled diamonds mark the posterior median of the start (left) and end (right) boundaries, with capped error bars giving their 95.4% HPDI (Table 1).

**Table 1. Bayesian phase-boundary estimates per technology, in cal BP.** For each technology the start and end of the dated phase were estimated as boundary parameters in a single uniform-phase model. Each cell gives the posterior median with the 95.4% HPDI in brackets. Duration is the difference between the start and end boundary medians.

| Technology | N | Start median [95% HDI] cal BP | End median [95% HDI] cal BP | Duration median years |
|---|---|---|---|---|
| Plant management | 35 | 11746 [12081–11453] | 7545 [7672–7271] | 4201 |
| Pottery | 24 | 6780 [7040–6570] | 4761 [4957–4495] | 2019 |
| Metallurgy | 7 | 3663 [4299–3409] | 2058 [2310–1406] | 1605 |
| Raised fields | 4 | 3471 [4819–3088] | 2085 [2664–680] | 1386 |

## 3.3. Asymmetric ENSO modulation of post-changepoint demographic activity

Late-Holocene fluctuations in palaeoclimate proxies were associated with the intensity of the radiocarbon record after the demographic change point identified by the double-exponential growth model (3575 cal BP, 95% HPDI 3811-3329 cal BP). We conducted the analysis aggregating calibrated dates into non-overlapping 100-year bins spanning 3811-500 cal BP (the older limit of the 95% HPDI of the change point and the approximate date of the contact with European groups, respectively), giving 34 successive 100-year bins. The radiocarbon dataset comprised 1374 dates falling within this window; 1355 passed the calibration filter (≥50% of the calibrated probability inside the window) and, after site-level binning (h = 100 years), yielded 1120 independent dated events. The summed calibrated probability falling in each bin was used as a continuous, SPD-derived event count (1120 events distributed across the 34 bins; bin counts ranged from 0 to 378). All models were fitted in NIMBLE using four independent chains of 100000 iterations (10000 burn-in, thinned by 2; 180000 retained samples). The Bayesian regressions converged satisfactorily in every case: the Gelman–Rubin statistic was $\hat{R} \approx 1.00$ for all parameters (upper confidence limit ≤ 1.01) and effective sample sizes exceeded 1400 for every parameter. Posterior estimates are reported as medians with 95% highest HPDI (Tables 2 and 3).

### 3.3.1. Pallcacocha XRF record for El Niño frequency

The ENSO covariate from lake Pallcacocha XRF record was associated with fluctuations of the human activity intensity, and both anomaly components had 95% HPDIs that excluded zero (Figures 4 and 5). The positive component, interpreted as proxy for more frequent El Niño events was positive ($\beta_+$ = 0.126; 95% HPDI 0.0081, 0.243), corresponding to a ~13% increase in the expected event rate per one standard deviation increase in El Niño frequency (rate ratio 1.13; 95% HPDI 1.01, 1.28). The negative component was substantially larger in magnitude and negative ($\beta_-$ = −0.87; 95% HPDI −1.146, −0.576): each one standard deviation strengthening of less frequent El Niño-like conditions was associated with roughly a halving of the event rate in the signal of human activity (rate ratio 0.42; 95% HPDI 0.32, 0.56, i.e. ≈58% fewer events). The background growth rate was small but well constrained and similar to the one approximated for $r_2$ in the double exponential growth model (r = 0.0016 $yr^{-1}$; 95% HPDI 0.0015, 0.0017), and the intercept was centred near zero (α = -0.045; 95% HPDI −0.331, 0.247).

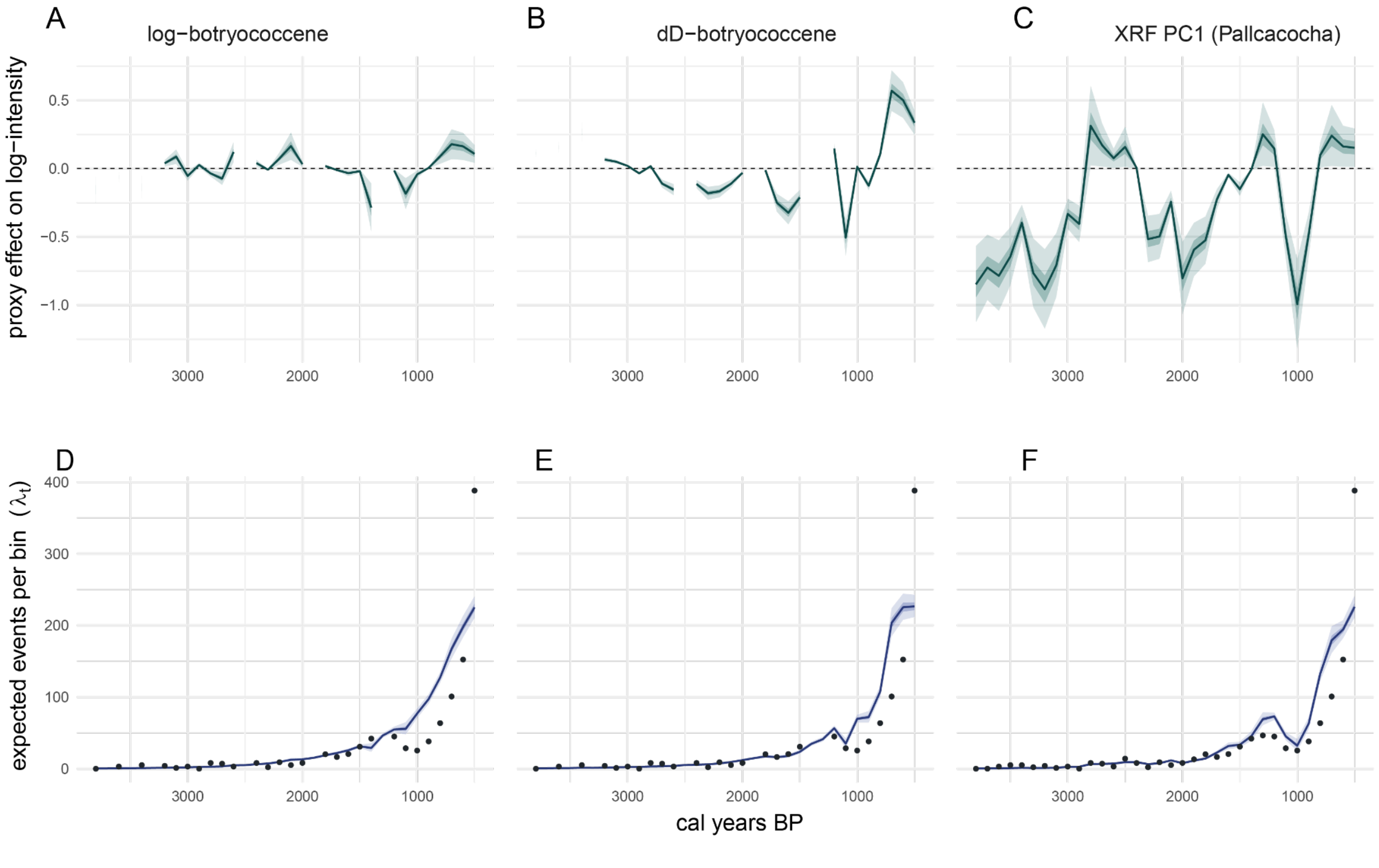

**Figure 4. Posterior per-bin proxy effects on demographic intensity and the resulting model fit to the radiocarbon record, for three palaeoclimate proxies over the 3811-500 cal BP window.** Columns correspond to the proxies tested: El Junco log-botryococcene (A, D), El Junco δD-botryococcene (B, E), and Pallcacocha XRF PC1 (C, F). Axes are shared (fixed) within each row. *Top row (A–C):* curves show only the proxy's contribution to log-intensity after baseline and trend are removed (the posterior of the proxy term at each bin). The solid line is the posterior median and the dark/light envelopes are the 50% and 95% HPDI credible bands. The dashed line at zero marks "no effect," so a band lying clearly above or below it indicates a credible increase or decrease in expected activity at that bin. The line is drawn only where proxy data exist and breaks at bins with no proxy measurement. Log-botryococcene proxy measurements are missing at 1300, 1900, 2500, 3300, 3500, and 3700 cal BP (6 of 34 bins), and δD-botryococcene at 1300, 1400, 1900, 2500, 3300, 3500, and 3700 cal BP (7 of 34 bins). *Bottom row (D–F):* points are the observed SPD event counts per bin included in the analysis. The solid line and shaded 50%/95% HPDI bands are the fitted intensity λ[t], drawn continuously across the full window: unlike the proxy effect, λ depends on the baseline and trend, which are global parameters defined at every age-bin, so the fitted curve is well defined even in bins lacking a proxy measurement. Age-bins without proxy measurements the proxy term is set to zero, so the fitted intensity there reflects baseline + trend only. A proxy is judged to matter when its top-row band departs from zero at the strong-proxy bins and its fitted line in the bottom row tracks the corresponding ups and downs of the SPD points.

**Table 2**. Posterior summaries for the bin-level intensity model with the Pallcacocha XRF PC1 ENSO covariate (Mark et al., 2022). Effect sizes are the exponentiated coefficients (multiplicative change in event rate per +1 SD of the covariate). α is the baseline log-intensity; *r* is the background growth rate ($yr^{-1}$); $\beta_+$ / $\beta_-$ are the effect of El Niño-like enhanced (frequency) / El Niño-like reduced (frequency) anomalies; ESS is the effective sample size; $\hat{R}$ is the Gelman–Rubin potential scale-reduction factor. The dash indicates that an exponentiated effect size is not interpreted for α and *r*.

| **Parameter** | **Median** | **Mean** | **SD** | **95% HPDI** | **exp(median) [95% HPDI]** | **ESS** | $\hat{R}$ |
|---|---|---|---|---|---|---|---|
| $\alpha$ | -0.045 | -0.044 | 0.148 | -0.331, 0.247 | – | 1841 | 1.0043 |
| $r$ | 0.0016 | 0.0016 | <0.0001 | 0.0015, 0.0017 | – | 2074 | 1.0045 |
| $\beta_+$ | 0.126 | 0.127 | 0.060 | 0.0081, 0.243 | 1.13 [1.01, 1.28] | 8225 | 1.0002 |
| *β*- | -0.867 | -0.866 | 0.145 | -1.146, -0.576 | 0.42 [0.32, 0.56] | 5396 | 1.001 |

### 3.3.2. El Junco biomarker proxies for El Niño frequency and amplitude

Both El Junco proxies showed a credible association with fluctuations in the human activity intensity, with 95% HPDIs for *β* that excluded zero (Figures 4 and 5). For log-botryococcene concentration the effect was positive (*β* = 0.108; 95% HPDI 0.043, 0.175), a 11% increase in the expected event rate per one standard deviation of the proxy (rate ratio 1.11; 95% HPDI 1.04, 1.19). For the δD of botryococcene the effect was negative (β = −0.236; 95% HPDI −0.298, −0.176), a 21% decrease in the event rate per one standard deviation (rate ratio 0.79; 95% HPDI 0.74, 0.84). Background growth rates were comparable to the double exponential growth model (r = 0.0018 for the log-botryococcene concentration and 0.0017 $yr^{-1}$ for the δD of botryococcene models), and the two intercepts were negative, reflecting the lower mean counts in the proxy-matched bins. Of the 34 bins, 28 carried log-botryococcene data and 27 carried δD data (the remaining data was excluded due to a lack of proxy observations). The majority of bins rested on a single proxy measurement (22 of 28 and 21 of 27, respectively). The opposing signs of the two coefficients are mutually consistent given that log-botryococcene concentration and δD track hydroclimate in opposite ways. Frequent El Niño-like conditions seem to enhance the human activity signal for pre-Columbian populations while reduced El Niño frequency and amplitude (intensity) seems to suppress human activity. Therefore, the enhanced and reduced ENSO conditions point to the same underlying relationship between the intensity of climate variability and patterns in the archaeological radiocarbon record during this interval.

**Table 3.** Posterior summaries for the bin-level intensity models fitted separately to the two El Junco proxies. Bins lacking proxy observations were excluded, leaving 28 of 34 bins for log-botryococcene concentration and 27 of 34 for δD of botryococcene; each model uses a single linear proxy coefficient *β*. Effect sizes are exponentiated coefficients (multiplicative change in event rate per one standard deviation of the proxy). α is the baseline log-intensity; *r* is the background growth rate ($yr^{-1}$); *β* is the effect of a one standard deviation change in the standardised proxy; ESS is the effective sample size; $\hat{R}$ is the Gelman–Rubin potential scale-reduction factor. The dash indicates that an exponentiated effect size is not interpreted for α and *r*.

| Parameter | Median | Mean | SD | 95% HPDI | exp(median) [95% HPDI] | ESS | $\hat{R}$ |
|---|---|---|---|---|---|---|---|
| **Log-botryococcene concentration** | | | | | | | |
| *α* | -0.792 | -0.79 | 0.139 | -1.058, -0.516 | – | 1711 | 1.0014 |
| *r* | 0.0018 | 0.0018 | 0.0001 | 0.0017, 0.0019 | – | 1694 | 1.0015 |
| *B* | 0.108 | 0.108 | 0.034 | 0.043, 0.175 | 1.11 [1.04, 1.19] | 6849 | 1.0006 |
| **δD of botryococcene** | | | | | | | |
| *α* | -0.586 | -0.584 | 0.155 | -0.892, -0.288 | – | 1444 | 1.006 |
| *r* | 0.0017 | 0.0017 | 0.0001 | 0.0016, 0.0018 | – | 1412 | 1.0057 |
| *β* | -0.236 | -0.236 | 0.032 | -0.298, -0.175 | 0.79 [0.74, 0.84] | 6038 | 1.0012 |

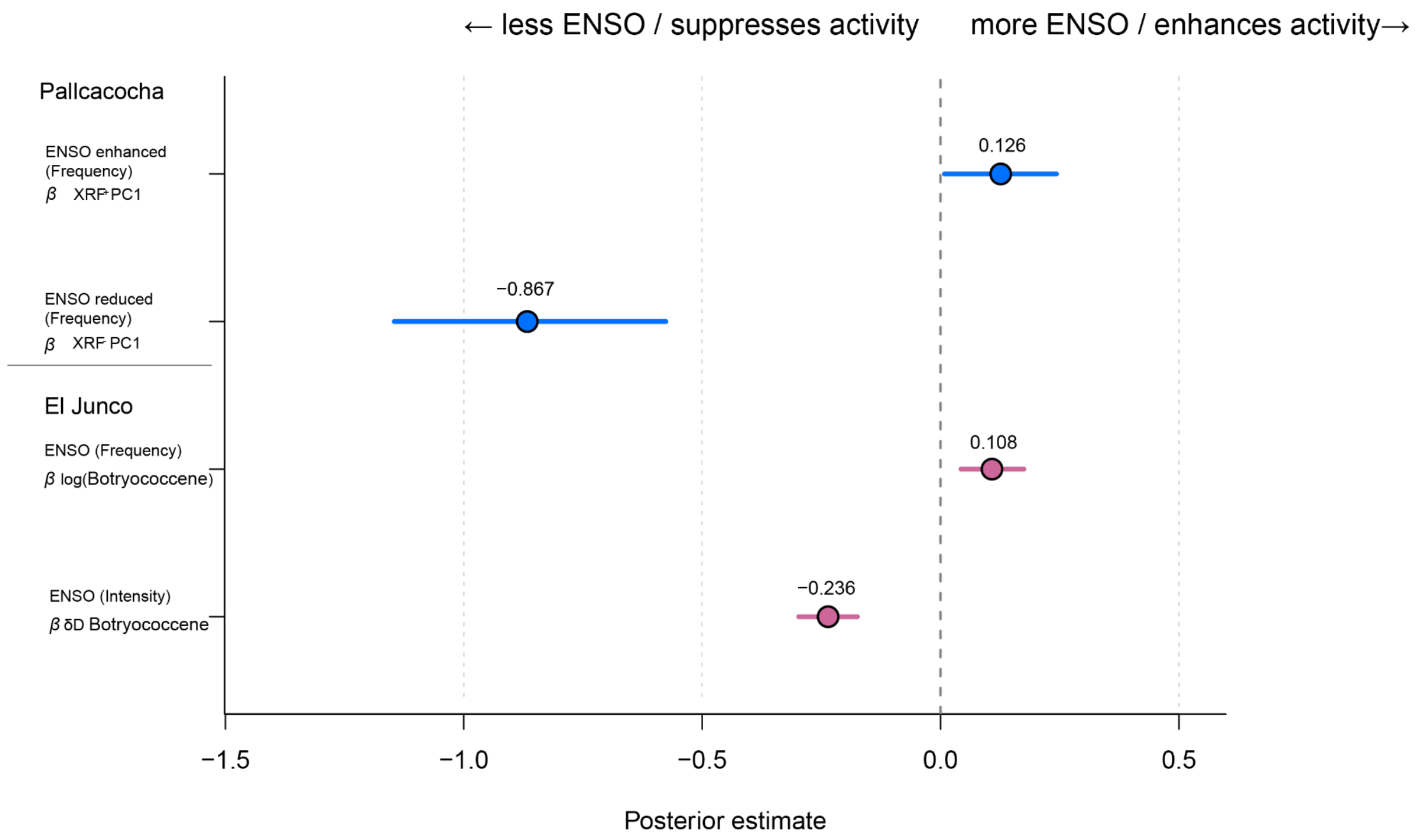

**Figure 5. Posterior effect sizes for the influence of ENSO-related palaeoclimate proxies (Pallcacocha and El Junco) on demographic intensity over the 3811-500 cal BP window.** Points show posterior medians and horizontal bars the 95% HPDI for the regression coefficients (*β*) of the bin-level log-linear intensity models. Values are on the log scale, so the vertical reference

line at 0 denotes no effect, and coefficients are interpreted as the multiplicative change in the SPD-derived event rate per one standard deviation of the (standardised) proxy. Negative coefficients (left) indicate suppression of human activity and positive coefficients (right) enhancement. Top (blue, Pallcacocha XRF PC1): the ENSO covariate is decomposed into positive and negative anomaly components, capturing changes in El Niño frequency, reflecting more events under enhanced El Niño frequency ($\beta^{+} = 0.126$; +13% per one standard deviation) and markedly fewer events as the reduced-El Niño (negative) anomaly strengthens ($\beta^{-} = -0.867$; −58% per one standard deviation). Bottom (pink, El Junco botryococcene biomarkers): a single linear coefficient per proxy was used for log-botryococcene concentration (an El Niño frequency indicator) and δD of botryococcene (an El Niño intensity indicator). These indicate slightly more events during enhanced El Niño frequency ($\beta = 0.108$; +11% per one standard deviation) and fewer events under reduced El Niño intensity (amplitude) ($\beta = -0.236$; −21% per one standard deviation). All four coefficients have 95% HPDIs excluding zero and all chains converged (Table 2 and 3).

## 3.4 Stability in demography and environment enables clusters of technologies

Metallurgy and raised fields technology overlap with the modelled demographic threshold. Metallurgy's posterior-median start boundary (≈3663 cal BP) effectively coincides with the changepoint posterior median (3575 cal BP) and lies inside its 95% HPDI (3811–3329 cal BP). The raised fields onset (≈3461 cal BP) also falls inside the changepoint 95% HPDI, immediately following the modelled threshold rather than preceding it. Both technologies therefore appear at, or just after, the transition from the slow to the faster demographic growth regime at a demographic plateau without positive or negative deviations of the expected growth rate identified in the double exponential model (Figure 6D).

Within the chronological window (≈3700–2000 cal BP) where metallurgy and raised fields emerged, the three climate proxies reveal contrasting temporal dynamics. The El Niño intensity proxy (δD, Figure 6 Panel B) is comparatively stable, oscillating within a narrow band (~ −255 to −265‰) with no large excursions until well after the windows of metallurgy and raised fields emergence close (the marked shift toward lower intensity occurs near ~1200 cal BP). This suggests a stable environment. The two frequency proxies, however, are not stable but their variability is less extreme than after the 2000 BP mark. XRF PC1 (Panel A) shows increasingly large positive excursions (more-frequent-El-Niño) developing from 1000 BP, while log-botryococcene (Panel C) swings repeatedly between ca. 2 and 4.7 mg/g with no sustained plateau indicating variable ENSO frequency with less extremes than after the 2000 BP mark. Therefore, the environmental backdrop against which we see technological emergence is one of stable El Niño intensity but variable El Niño frequency.

During this chronological window, the human-activity proxy remains largely stable. At the onset of this phase, the SPD remains low and just beginning its post-threshold rise, meaning the technologies emerged during a period of steady demographic change. Through the core of the window the observed SPD broadly tracks the posterior-predictive envelope. However, toward the end of the metallurgy and raised field intervals—approximately 2200–2000 cal BP—the observed SPD rises above the model envelope. This positive deviation (in red), indicates that human activity exceeded baseline demographic expectation precisely as these technological emergence phases drew to a close.

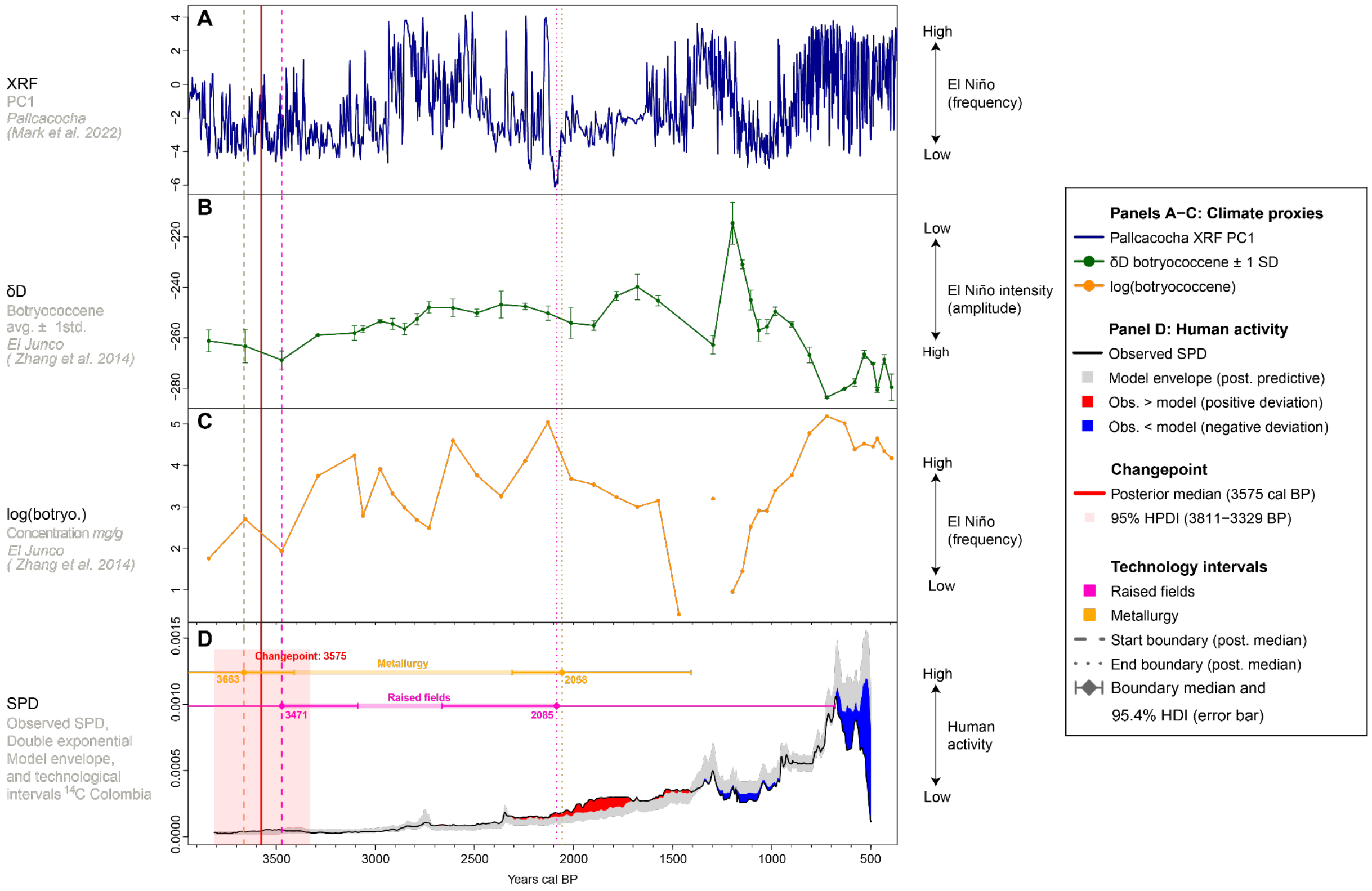

**Figure 6. Multi-proxy comparison of late-Holocene ENSO-related climate, technological phases, and human activity in Colombia over the 3811-500 cal BP window**. (A) Pallcacocha XRF PC1 (Mark et al., 2022), a proxy for El Niño frequency. (B) δD of botryococcene (mean ± one standard deviation) from El Junco, Galápagos (Zhang et al., 2014), a proxy for El Niño intensity (amplitude); note the inverted axis, with more negative values indicating higher intensity. (C) log-botryococcene concentration (mg/g) from El Junco (Zhang et al., 2014), a second El Niño frequency proxy. (D) Human activity: the observed summed probability distribution of calibrated $^{14}C$ dates (SPD, black) against the posterior-predictive envelope of the fitted double-exponential growth model (grey); red and blue mark intervals where the observed SPD lies above (positive deviation) or below (negative deviation) the model envelope. The vertical red line and pink band give the posterior median (3575 cal BP) and 95% HPDI (3811–3329 cal BP) of the demographic change point. Horizontal bars show the start and end estimations of the emergence window. Filled

diamonds mark the posterior median of the start (left) and end (right) boundaries, with capped error bars giving the 95.4% HPDI (Table 1) for metallurgy (≈3663–2058 cal BP) and raised fields (≈3471–2085 cal BP), with dashed and dotted lines marking the posterior-median start and end boundaries.

# 4. Discussion

## 4.1. Technological emergence and stability

The use of four analytical components within a single temporal frame allows us to evaluate a set of key questions about technological advances in human societies and climate influence.

The first question concerns whether the technology intervals appear clustered in time, and how such clustering sits relative to the demographic threshold. The technology intervals do not describe one continuous episode of co-occurring innovation across the Holocene but they highlight two contrasting patterns. Plant management and pottery emergence both occurred within the prolonged, near-stationary growth phase, separated by a sufficient interval (~800 years) to prevent their chronological ranges from overlapping. These technologies probably behaved as isolated, slow-diffusing acquisitions, mirroring the pattern of a demographic transition postdating the adoption of domesticates documented in Europe (Shennan et al., 2013) and southeastern South America (de Souza and Riris, 2021). In contrast, metallurgy and raised-field construction formed the only temporal pairing in the record—a pairing anchored directly to the demographic changepoint rather than preceding it. The modelled onset of metallurgy and raised field construction fall inside the credible interval of the changepoint. This technological single cluster coincides with the transition from slow to accelerated signal of pre-Columbian human activity. Below that threshold, the record reflects the expectations of the collective-computation model: innovations that are not lost, but they diffuse slowly and not synchronously. The clustered pair appears only once the threshold is crossed.

We interpret the post-changepoint acceleration as a rise in the probability of intensified human activity rather than as a direct estimate of population size. Radiocarbon summed probabilities aggregate signals from population growth, expanding interaction networks, and increasing per-capita activity, and they do not by themselves identify which of these mechanisms drives a given acceleration (Brown, 2017; Crema and Shoda, 2021; Timpson et al., 2014; Williams, 2012). Independent regional evidence indicates that the most pronounced rise in residential population in Colombia occur later, as early as ~2000 cal BP in Alto Magdalena (Drennan et al., 2026, 2018) and ~1000 cal BP in the Eastern Cordillera (Drennan et al., 2026; Fajardo and Argüello, 2024;

Langebaek, 1995; Hope Henderson, 2017), the Cauca Valley (Giraldo, 2014), and the Sierra Nevada de Santa Marta (Rodríguez, 2026; Ruiz et al., 2022). This pattern suggests that the 3575 cal BP change point reflects an organisational change—increased connectivity or activity intensity per occupation—rather than a simple population growth. Alternatively, this signal may partly reflect a shift in archaeological visibility rather than past behaviour. Longer occupations in single locations and more intensive landscape intervention also produce a more detectable record, an effect reinforced by lower taphonomic loss expected from recent deposits. The characteristics of the sites falling within the HPDI of the changepoint favour a behavioural reading over one driven by preservation or detectability. Sites like Aguazuque (Correal, 1990) and Chía 1 (Ardila, 1984) in the Sabana de Bogotá; Barlovento (Reichel-Dolmatoff, 1955), Monsú (Reichel-Dolmatoff, 1985), San Jacinto II (Oyuela-Caycedo and Bonzani, 2005) and Malambo (Angulo, 1981) on the Caribbean coast; La Mikaela (Cano, 2004; Dickau et al., 2015) in the Cauca valley; and La Maporita (Romero Picón, 2009) in the Eastern Lowlands—did not go on to become long-lived, intensively occupied centres that anchored the demographic growth later documented in regional settlement patterns during the emergence of higher levels of sociopolitical complexity. Were the changepoint largely an artefact of visibility or differential preservation, its signal should be carried disproportionately by the sites that became most salient and continuous through time; instead it rests on sites that did not subsequently intensify, suggesting a genuine organisational shift at ~3575 cal BP rather than a bias of the record.

The second question concerns whether the identified technological cluster corresponds to environmental stress, expressed as demographic or climatic variability. Neither the onset of metallurgy or of raised fields aligns with a peak in climate variability or with a departure of the observed SPD from its expected trajectory. The most interesting result is that, across the interval bracketing such onsets, El Niño intensity is comparatively flat and the observed record for pre-Columbian human activity tracks the posterior-predictive envelope without sustained positive or negative excursion. The marked SPD departures from prediction and the largest proxy excursions develop only later—well after the window for clustered technological emergence opens and toward its close. El Niño frequency remains variable while its intensity is stable, so the evidence suggests a qualified rather than absolute "stable-window" reading. However, the evidence does not support that innovation would arise with climate variability maxima or at demographic anomalies. The two technologies, actually, emerge on a demographic plateau under stable ENSO intensity, with the demographic threshold itself—not any episode of environmental stress—providing the temporal anchor.

Our claim concerns the basin-scale trend in ENSO intensity, which remains comparatively stable across the window for clustered technological emergence. We are not claiming spatial homogeneity of climate within Colombia. The regional precipitation dipole across the window of interest, with the Eastern Cordillera wetter than today (~4200–2600 cal BP) and western Colombia anomalously dry (~3700–1200 cal BP) documented in previous work (Velez et al., 2026) is compatible with our interpretation. This geographic pattern may help account not for when the technological cluster emerges but for where and in what form it does. Raised fields appear first in the Eastern Cordillera at 3260 Cal BP (Boada, 2018), precisely the region reconstructed as anomalously wet. Canals appear in the Caribbean lowlands later as early as 2982 cal BP (Plazas et al., 1993). This onset is consistent with a local adaptive response to excess precipitation in the Eastern Cordillera with later acquisition in other geographical regions. Metallurgy appears first between 3663 and 2058 cal BP on the Pacific coast (Patiño, 1997) and possibly in the Caribbean lowlands (Plazas, 1998)—catchments draining the Western Colombia, where conditions were drier. Reduced flow under drier conditions could have exposed gold-bearing river sediments and increased the visibility and accessibility of placer deposits. We stress that our data cannot adjudicate among these mechanisms or establish causation. What the spatial coincidence does indicate is that the two technologies need not be treated, by default, as a single undifferentiated cluster. It raises the question of whether their near-contemporaneous emergence was linked at all, and if so, through what mechanisms. We favour an explanation grounded in social connectivity and inter-regional exchange over a climate-driven one: the demographic threshold, not a climatic episode, supplies the temporal anchor, and the regional dipole at most modulates the local form each technology took. Whether the cluster reflects genuine connection or convergent cultural evolution under a shared demographic ceiling remains open, but it is the structure of interaction between regions, rather than climate forcing, that we see as the more promising line to pursue.

We did not formally model climate modulation for the period where plant management and pottery emerged, since the regression was restricted to the post-changepoint interval. The qualitative picture, however, points to a similar dynamic. Through both windows the observed SPD remains within the model's posterior-predictive envelope, so there is no demographic departure to account for, and the independent ENSO reconstruction indicates comparatively subdued variability over the relevant intervals with reduced El Niño activity in the Middle Holocene and only moderate activity in the Early Holocene (Carré et al., 2014; Mark et al., 2022). This is consistent with the emergence of such technologies as part of mobile, broad-spectrum foragers (Aceituno, 2026) dynamics to buffer short-term climate fluctuations. Neither technology emerges during a period of pronounced environmental stress, similarly to what we see for the modeled metallurgy and raised fields emergence.

The third question asks how much of the post-changepoint activity record is attributable to climate once the underlying growth trend is removed. Estimating the marginal effect of each proxy on binned intensity returns a contribution that is genuine but modest and, more importantly, asymmetric. While a decrease in El Niño conditions depresses the activity signal—effectively halving it per standard-deviation change—more frequent or intense El Niño conditions yield only a negligible increase. Climate variability thus appears to have modulated the intensity of human activity rather than driving the demographic departures themselves, which emerge only after the introduction of new technologies. The quantitative apportionment reinforces this interpretation: ENSO affected how strongly the activity signal developed, particularly limiting its growth during periods with fewer El Niño-like conditions, but it was not the catalyst for the cluster of technological innovations.

When these results are taken together, align naturally with a collective-computation view of innovation (Arthur, 2009; Flack, 2017; Gabora, 2013; Vélez et al., 2023), where the rate of recombinatorial novelty depends jointly on the size of the interacting population and on the system's proximity to a critical regime.

ENSO shaped the conditions under which innovation could happen rather than triggering it directly, by influencing both how much time a population could devote to the collective work of generating innovations and how predictable the environment was for organising daily life. When El Niño-like conditions were reduced, rainfall changes in Colombia may have intensified, altering environmental patterns even across large ecological regions whose earlier conditions had been stable or only moderately variable (Velez et al., 2026). This loss of dependable environmental prediction weakened the interacting population on which sustained innovation depends, and human activity declined. Greater climate variability also failed to stimulate innovation: because the system (society) was already at the critical threshold between stability and disorder, further disruption induced stochastic unpredictability rather than adaptive, useful change. The corollary is that the changepoint marks less a simple increase in head-count than an organisational transition—greater connectivity or higher per-occupation activity—above which the clustered emergence of technologies became possible. Effective-population growth is necessary but not sufficient; comparative stability in both the climate fluctuations and the demographic dimensions appears to be the additional condition under which clustered innovation occurs.

Greater connectivity through intra-group cohesion is one possible mechanism for population stability when conditions require coordinating and managing resources above the kin level, as documented during the Medieval Climate Anomaly (~1000–700 cal BP) in southern Brazil (Riris and de Souza, 2021) and in Colombia during this same period when social complexity appears to have expanded (Drennan, 1991; Fajardo and Argüello, 2024; Oyuela-Caycedo, 2008; Rodríguez, 2026). We would extend this argument: connectivity may emerge earlier and enable the development and sharing of new cultural traits, rather than functioning only as a mechanism for ensuring stability.

These observations must be assessed against a growing body of empirical work that integrates summed radiocarbon distributions with high-resolution palaeoclimatic records. In Amazonia and circum-Amazonia, for instance, sustained demographic downturns appear in intervals of high climatic variability (Riris and Arroyo-Kalin, 2019), a counter image of the relationship inferred here, where innovation does not emerge in variability maxima. The strength and direction of ENSO's modulation on human activity can vary across regions, showing patterns different from as the one documented here. On the coast of Perú, for instance, mid-Holocene weakening or absence of ENSO appears to have benefited coastal populations. (Riris, 2018). Globally, the demographic signature of human activity after the occurrence of external or internal disturbances is one of resistance and recovery (Riris et al., 2024). Complementing this picture, innovation clustering aligns in our record with the demographic threshold under relative stable conditions rather than with disturbance episodes. The case for the current Colombian territory adds a tropical and technology-focused example to such previous work and one in which the clearest signal is the alignment of innovation with a demographic threshold under stability rather than with stress.

## 4.2. Caveats and limitations

While the inferences proposed in this paper provide a compelling framework, they must be interpreted within the context of several inherent data and chronological constraints. Identifying these boundaries is essential for a transparent assessment of the record; however, they do not compromise the robustness of our model or negate the fundamental patterns observed in the results.

The directly dated technology sample is small and plant management contexts have a higher presence. This limits comparison among technological classes and the metallurgy and raised-field intervals—which carry the clustering argument—rest on fewer determinations. Because all of these intervals are estimates, earlier emergence cannot be excluded. However, for metallurgy and raised fields a marginally earlier onset would still fall within the changepoint window, so the alignment we discuss is robust in direction even if the precise boundaries might shift.

The climate-coupling interpretations apply only to the post-changepoint interval and are therefore directly relevant to metallurgy and raised-field agriculture. The long term accumulation of plant-management knowledge predates this threshold and falls outside the tested interval, preventing an assessment of its relationship with climate variability.

The two proxy systems are regional rather than locally associated to where technology appears. ENSO variability is captured at the level of current Colombia territory without direct hydroclimatic evidence at the archaeological sites. Furthermore, El Junco series resolves climate at a coarser, near-centennial level than the quasi-continuous XRF record of sediment from laguna Pallcacocha, so apparent intensity "stability" may partially reflect sampling density. The asymmetric coding of the Pallcacocha predictor and the single-slope coding of the El Junco proxies also test slightly different hypotheses, and the directional agreement among them should be read with that difference in mind. Finally, the alignment is correlational and the SPD aggregates population growth, network expansion, and per-capita activity that the radiocarbon record alone cannot separate; the qualitative conclusions are anchored by the convergence of three independent proxies, but they are not a formal causal test.

## 4.3. Future directions

To overcome data resolution and limitation on sampling, and to refine the scope of our conclusions, future research should prioritize four distinct lines of work. First, augmenting the directly dated technology sample—especially with determinations spanning the changepoint window—would tighten the intervals and allow sub-regional comparison. Second, incorporating local high-resolution palaeoclimate archives, such as lake cores and speleothem records, would test whether the ENSO activity coupling operates through local conditions, allowing for the regional proxies to be cross-validated against site-proximal signals. Third, coupling the radiocarbon record to independent demographic proxies—settlement survey, ancient DNA, or stable-isotope dietary series—would help disentangle the population, network, and individual components that the SPD conflates and clarify what the changepoint represents. Fourth, encoding the climate-stress and collective-computation hypotheses as competing generative models and adjudicating them with Bayes factors or cross-validated information criteria would convert the present weight-of-evidence assessment into a properly inferential one. Finally extending the whole framework across a larger geographic region would distinguish a general mechanism from a Colombia-specific narrative.

# 5. Conclusions

Whether transformative technologies emerge in response to climate variability or as a function of population-scale capacity has rarely been tested jointly, against the same record and within one inferential framework, in a tropical setting. Treating technological emergence as a collective-computation problem, we used the Colombian Holocene record to ask whether innovation tracks climate variability, demographic departures, or windows of relative stability.

The only technological cluster—metallurgy with raised field construction—resolves at an organisational changepoint in human activity intensity, while earlier acquisitions (plant management and pottery production) behave as slow, distributed processes that the threshold neither triggers nor suppresses. That cluster does not coincide with climatic turbulence or with demographic anomalies, which arise only later; and where climate does register, its effect on activity is modest and one-sided, depressing the signal during arid, reduced-El Niño intervals rather than driving innovation during stress. Climate, in short, regulates the substrate; it does not author the cascade.

These findings favour a regulator–substrate coupling over a strong stress account: environmental stability of a well connected population network may constitute an important condition for clustered innovation. Across the Colombian Holocene record, the deepest transformations did not arrive under strong variable conditions, but when a population had grown connected enough to recombine their knowledge and stable enough to keep doing so.

# Declaration of competing interest

The authors declare that they have no known competing financial interests or personal relationships that could have appeared to influence the work reported in this paper.

## CRediT authorship contribution statement

Conceptualization: S.F., P.R.; Data curation: S.F., N.A., A.J.H., L.F.N., R.C., N.L., C.R.; Formal analysis: S.F.; Funding acquisition: M.M.; Investigation: S.F.; Methodology: S.F., P.R.; Project administration: S.F., M.M.; Software: S.F.; Resources: S.F.; Supervision: S.F., M.M.; Visualization: S.F.; Writing – original draft: S.F., A.J.H, L.F.N., R.C.; Writing – review & editing: S.F., P.R., M.V., N.A., A.J.H., L.F.N., R.P. N.L., C.R., J.R.P. , M.M.

## Acknowledgments

This research arises from the Mapping the Archaeological Pre-Columbian Heritage of South America - MAPHSA project funded by the Arcadia Fund, a charitable fund founded by Lisbet Rausing and Peter Bakdwin. PR is funded by an Arts and Humanities Research Council grant (AH/X002217/1). We thank Lina María Campos Quintero, Danna Díaz Chacón, Roberto Lleras Pérez, Juan Camilo Vásquez, and Jasmine Vieri for providing information to support the data curation and validation process. We are also grateful to all the researchers who generated the radiocarbon dated samples that make up this dataset, whose original work made this study possible.

## Data availability

All radiocarbon data, processing scripts, and computational outputs are openly released and actively maintained in the GitHub repository (COL14). A versioned release of all code, data, and outputs is permanently archived in a public Zenodo repository (Fajardo et al., 2026 dataset DOI: 10.5281/zenodo.21084728). Both the GitHub repository and the Zenodo archive are cross-linked to the MAPHSA database. The published environmental proxy records used in this study were obtained from their original sources. The Laguna Pallcacocha XRF record is available from the NOAA World Data Service for Paleoclimatology (Mark et al., 2022 dataset DOI: 10.25921/41b1-w758; EPSL, DOI: 10.1016/j.epsl.2022.117657). El Junco biomarkers record is available from its cited primary source (Zhang et al., 2014).